\documentclass{amsart}

\usepackage{amssymb,latexsym,amsfonts,amsmath}
\usepackage{graphicx}
\usepackage{eso-pic}
\usepackage{epsfig}
\usepackage{graphicx}
\usepackage{color}
\usepackage{algorithm}
\usepackage{algorithmic}
\usepackage{subfigure}
\PassOptionsToPackage{numbers, compress}{natbib}

\usepackage[utf8]{inputenc} 
\usepackage[T1]{fontenc}    
\usepackage{hyperref}       
\usepackage{url}

\usepackage{amssymb,latexsym,amsfonts,amsmath}
\usepackage{graphicx}
\usepackage{amsthm}
\usepackage{amsmath}

\newtheorem{theorem}{Theorem}[section]

\newtheorem{prob}[theorem]{Problem}

\theoremstyle{definition}
\newtheorem{defn}[theorem]{Definition}

\theoremstyle{remark}

\numberwithin{equation}{section}

\newcommand{\R}{{\mathbb{R}}}

\newcommand{\MSE}{\mathsf{MSE}}

\newcommand{\Sys}{\mathfrak{S}}
\newcommand{\Xx}{\mathcal{X}}

\newcommand{\Relu}{\mathsf{ReLU}}

\DeclareSymbolFont{bbold}{U}{bbold}{m}{n}
\DeclareSymbolFontAlphabet{\mathbbold}{bbold}

\newcommand{\real}{\mathbb{R}}

\begin{document}

\begin{abstract}
This report presents a neurosymbolic framework for safety verification and control synthesis in high-dimensional \emph{monotone} dynamical systems without relying on explicit models or conservative Lipschitz bounds.
The approach combines the expressiveness of neural networks with the rigor of symbolic reasoning via \emph{barrier certificates}, functional analogs of inductive invariants that formally guarantee safety.
Prior data-driven methods often treat dynamics as black-box models, relying on dense state-space discretization or Lipschitz overapproximations, leading to exponential sample complexity. 
In contrast, monotonicity---a pervasive structural property in many real-world systems---provides a symbolic scaffold that simplifies both learning and verification. 
Exploiting order preservation reduces verification to localized boundary checks, transforming a high-dimensional problem into a tractable, low-dimensional one.
Barrier certificates are synthesized using \emph{monotone neural networks}---architectures with embedded monotonicity constraints---trained via gradient-based optimization guided by barrier conditions. 
This enables scalable, formally sound verification directly from simulation data, bridging black-box learning and formal guarantees within a unified neurosymbolic framework.
\end{abstract}

\title[Monotone Neural Barrier Certificates]{Monotone Neural Barrier Certificates}

\author[Saber Jafarpour]{Saber Jafarpour$^1$} 
\author[Alireza Nadali]{Alireza Nadali$^1$} 
\author[Ashutosh Trivedi]{Ashutosh Trivedi$^1$} 
\author[Majid Zamani]{Majid Zamani$^1$}
\address{$^1$}
\email{\{saber.jafarpour, a\_nadali, ashutosh.trivedi, majid.zamani\}@colorado.edu}
\urladdr{https://saberjafarpour.github.io}

\maketitle

\section{Introduction}

Discrete-time dynamical systems offer a foundational framework for modeling and controlling a broad spectrum of digitally controlled autonomous systems, including autonomous vehicles, implantable medical devices, biological networks, and industrial process management. 
As these systems constitute a critical component of our safety infrastructure, high-assurance formal reasoning is essential to ensure their safe operation. A principled and efficient approach to providing such safety guarantees involves synthesizing {\it barrier certificates}~\cite{prajna2006barrier,ADA-SC-ME-GN-KS-PT:19}---a functional analog of inductive invariants---whose existence serves as a formal proof of safety~\cite{AP-AR:04}.
Although finding such certificates is inherently computationally challenging, recent neuro-symbolic approaches have shown considerable promise through the use of {\it neural barrier certificates}~\cite{chang2019neural,zhao2020synthesizing,zhang2024exact,robey2020learning}. 
While these neural certificates provide substantial flexibility in searching for safety guarantees, we posit that scalability and sample efficiency can be further improved in specialized control applications by identifying and leveraging structural properties of the underlying system.
In this work, we focus on a subclass of discrete-time dynamical systems known as \emph{monotone systems}~\cite{angeli2003monotone,HLS:95}, which frequently arise in domains such as queuing networks~\cite{farina2011positive}, population dynamics~\cite{leenheer2004predator}, biological systems~\cite{DA:13}, and chemical networks~\cite{EDS:07}. 
By leveraging their structural characteristics, we synthesize barrier certificates represented as \emph{monotone} neural networks, which we refer to as \emph{monotone neural barrier certificates}.

\vspace{0.3em}\noindent\textbf{Monotone Systems.} Monotone dynamical systems~\cite{HLS:95,angeli2003monotone} are characterized by the property that their trajectories preserve a natural partial order induced on their state set.
Examples of monotone systems arise in domains such as biochemical reactions in synthetic biology~\cite{EDS:07}, frequency regulation in power grids, congestion control in transportation networks~\cite{SC-MA:15,GC-EL-KS:15}, population dynamics in ecology, and compartmental models in epidemiology~\cite{angeli2003monotone}. 
Related concepts have also been studied in other domains, including positive linear systems~\cite{farina2011positive}, well-structured transition systems~\cite{FINKEL200163}, and well-behaved transition systems~\cite{Blondin2016WellBT}. 
It is well known that monotone systems exhibit highly ordered transient and asymptotic behaviors~\cite{HLS:95}, making them particularly well-suited for data-driven methods, as their inherent structure helps mitigate the challenges posed by limited or noisy data.
While monotonicity has informed control and stability analysis~\cite{AR:15,SC:19}, its potential for learning-based safety verification remains largely untapped.

\vspace{0.3em}\noindent\textbf{Barrier Certificates (A Proof of Safety).}
In safety verification, the central object of interest is the \textit{reachable set}, which comprises all states the system can reach from a given set of \textit{initial states} under a fixed control policy.  
Safety can be certified by showing that this set does not intersect the set of harmful or \textit{unsafe states}.  
However, computing the exact reachable set is generally intractable for nonlinear or high-dimensional systems.  
A common alternative is to construct an \textit{inductive state-invariant}, i.e., a set that contains the initial states and is \emph{closed} under the system’s transitions.  
Barrier certificates~\cite{prajna2006barrier} offer one such approach by defining a real-valued function over the state space whose level sets form a ``barrier'' separating safe and unsafe regions.  
Specifically, the barrier function must:  1) take values less than or equal to a constant over the initial states, 2) and strictly greater than that constant over unsafe states; and  3) decrease along system trajectories outside the unsafe set.
A control barrier certificate (CBC) generalizes the idea of barrier certificate by requiring the existence of an action satisfying the conditions for a barrier certificate.

Despite their intuitive appeal and ease of verification, complexity of finding CBCs remains a significant hurdle in their widespread adoption.
Traditional methods for synthesizing CBCs rely on optimization-based formulations such as Sum-of-squares (SOS) programming~\cite{AP-SJ:05,kong2013exponential}. 
SOS is a popular approach that is applicable to systems with polynomial dynamics; however, it quickly becomes intractable due to the growth of semidefinite constraints with system dimension and polynomial degree.
Linear programming (LP) methods, while more tractable, are typically limited to linear or affine certificates and cannot capture the rich nonlinear boundaries present in many real-world systems~\cite{SS-XC-EA:13,YZ-etal:16}. 
Recent progress in deep learning has enabled the use of neural networks to be trained as barrier certificates~\cite{anand2023formally,lindemann2024learning,dawson2022safe,liu2023safe,zhao2020synthesizing,Rus2023,zhang2024exact}.
These so-called neural control barrier certificates (NCBCs) represent both the barrier function and controller as neural networks.
While expressive, these approaches often require full knowledge of the system dynamics (to verify correctness using constraint satisfaction tools such as Z3~\cite{de2008z3}) or rely on bounding {\it Lipschitz constants} (which entails exponential sample complexity), limiting scalability and applicability to black-box systems~\cite{anand2023formally,zhang2024exact}. 
Moreover, these approaches treat the system as unstructured, ignoring properties like monotonicity that could drastically reduce the number of required samples and constraints.

\vspace{0.3em}\noindent\textbf{Monotone Neural Control-Barrier Certificates.}
We introduce an efficient scheme for synthesizing barrier certificates specifically tailored to (unknown) \textit{ monotone systems}.
By exploiting monotonicity, we restrict the satisfaction of safety conditions to critical boundary slices of the state space—namely, the upper boundary of the initial set and the lower boundary of the unsafe set. 
This reduces the verification burden from the entire state space to a carefully selected subset---without compromising formal guarantees---and enables the use of gradient-based optimization to learn barrier certificates using simulation data alone.
Our method avoids the need for explicit models, Lipschitz bounds, or semidefinite relaxations.
This makes it uniquely suited for safety verification in large-scale, nonlinear systems where monotonicity arises naturally from physical or structural properties.

\section{Preliminaries and Mathematical Foundations}
 
\paragraph{\textbf{Notations.}} We denote the set of real, non-negative real, and positive real numbers by \( \R \), \( \R_{\geq 0} \), and \( \R_{>0} \), respectively. For sets \( A \) and \( B \), we use \( A \setminus B \) for set difference and \( A \times B \) for Cartesian product, and write \( |A| \) for the cardinality of \( A \). 
Given a vector \( v = (v_1,\ldots,v_n)^\top \in \R^n \), its Euclidean norm is \( \|v\| = \sqrt{\sum_{i=1}^n v_i^2} \). The rectified linear unit is defined by \( \Relu(x) := \max(x,0) \), and for vectors \( x, y \in \R^n \), the mean squared error is \( \MSE(x, y) := \frac{1}{n} \sum_{i=1}^n (x_i - y_i)^2 \). We denote the \( i \)th component of a vector \( x \) by \( x_i \), and write \( x \le y \) to indicate element-wise inequality.
A set \( \mathcal{S} \subseteq \R^n \) is \emph{lower closed} if \( x \in \mathcal{S} \) and \( y \le x \) imply \( y \in \mathcal{S} \); it is \emph{upper closed} if \( x \in \mathcal{S} \) and \( y \ge x \) imply \( y \in \mathcal{S} \). A hyper-rectangle is denoted by \( [\underline{x}, \overline{x}] := \{ x \in \R^n \mid \underline{x} \le x \le \overline{x} \} \). Given a set $A\subseteq \real^n$, $x\in \real^n$ is a maximal (resp. minimal) point of $A$, if there does not exists any $y\in A$ such that $x\le y$ (resp. $y\le x$).

\subsection{Safety of Dynamical Systems}
%
In this section, we provide a formal definition of discrete-time dynamical systems and establish a rigorous notion of safety. We then present two central safety objectives. 

\begin{defn}[Discrete-time dynamical system]
    \label{sys}
A discrete-time control system (dtDS) is a tuple $\Sys :=(\Xx,\Xx_0,U,f)$, where $\Xx \subseteq \R^n$ represents the state set, $\Xx_0 \subseteq \Xx$ is the initial state set, $U \subseteq \R^m $ is the set of inputs, and $f:\Xx \times U \to \Xx$ is the state transition map.
\end{defn}
Given an input sequence $\{u_t\}_{t=0}^{\infty} \in U$, the evolution of $\Sys$ under this input is described by:
\begin{equation}\label{eq:dynamics}
    x_{t+1} = f(x_t, u_t), \qquad \text{for all } t \ge 0.
\end{equation}
The sequence $\{x_t\}_{t=0}^{\infty}$ defined by~\eqref{eq:dynamics} is called a \textit{trajectory} of $\Sys$ for the input $\{u_t\}_{t=0}^{\infty}$. 
Given a dtDS $\Sys = (\Xx, \Xx_0, U, f)$, a set of unsafe states $\Xx_u \subseteq \Xx$, and an input sequence $\{u_t\}_{t=0}^{\infty}$, we say that $\Sys$ is \textit{safe} with respect to $\Xx_u$ for this input sequence if all trajectories starting from initial states in $\Xx_0$ remain outside of $\Xx_u$. That is, for every $x_0 \in \Xx_0$, the corresponding trajectory $\{x_t\}_{t=0}^{\infty}$ for the input sequence $\{u_t\}_{t=0}^{\infty}$ starting from $x(0) = x_0$, satisfies
\begin{align*}
    \{x_t\}_{t=0}^{\infty} \cap \Xx_u = \emptyset. 
\end{align*}
In dynamical systems, how inputs are managed fundamentally shapes safety objectives. Hence, two core challenges arise: the safety verification problem and the safe control synthesis problem.
\paragraph{\textbf{Safety Verification Problem.}} 
In safety verification, inputs are viewed as disturbances, and the goal is to guarantee system safety under \textit{all possible} input sequences $\{u_t\}_{t=0}^{\infty} \in U$. To this end, \textit{robust barrier certificates} have been widely employed as an effective tool for safety verification~\cite{prajna}.

\begin{defn}[Robust barrier certificates]
\label{cbc_def}
Consider a dtDS \sloppy  $\Sys = (\Xx,\Xx_0,U,f)$ with a set of unsafe states $\Xx_u\subseteq \Xx$. A function $B:\Xx\rightarrow \R$ is called a robust barrier certificate (RBC) for $\Sys$ with respect to the unsafe set $\Xx_u \subseteq \Xx$ if there exists $ \gamma,\delta,\eta \in \R_{\geq 0}$ such that $\gamma\leq \delta\le \eta$, and: 
\begin{align}
    &B(x) \leq \gamma, \qquad\text{ for all } x \in \Xx_0,\label{eq_barr_11}\\
    &B(x)  > \eta,   \qquad\text{ for all } x \in \Xx_u, \text{ and } \label{eq_barr_21} \\
   & B(x) \leq \delta \implies \mathrm{sup}_{u\in U}\{B(f(x,u))\}\leq \delta.  \label{eq_barr_31}
\end{align} 
\end{defn}
The following theorem formulates the safety verification problem as a search for RBCs~\cite{prajna}. 
\begin{theorem}[Safety verification via robust barrier certificates]\label{thm:safebarrier}
\label{cbcthm}
Consider a dtDS $\Sys = (\Xx, \Xx_0, U, f)$ and a set of unsafe states $\Xx_u \subseteq \Xx$. Suppose that there exists a robust barrier certificate satisfying conditions~(\ref{eq_barr_11})-(\ref{eq_barr_31}). For any input sequence $\{u_t\}_{t=0}^{\infty} \in U$, the system $\Sys$ is safe with respect to $\Xx_u$. 
\end{theorem}

\paragraph{\textbf{Safe Control Synthesis Problem.}}
In safe control synthesis, inputs are treated as control policies, aiming to \textit{synthesize} a controller $\{u_t\}_{t=0}^{\infty}$ that ensures system safety. \textit{Control barrier certificates} have been widely adopted as an effective tool for synthesizing such controllers~\cite{ADA-XX-JWG-PT:17}.

\begin{defn}[Control barrier certificates]
\label{cbc_def2}
Consider a dtDS \sloppy  $\Sys = (\Xx,\Xx_0,U,f)$ with a set of unsafe states $\Xx_u\subseteq \Xx$. A function $B:\Xx\rightarrow \R$ is called a Control Barrier Certificate (CBC) for $\Sys$ with respect to the unsafe set $\Xx_u \subseteq \Xx$ if there exists $ \gamma,\delta,\eta \in \R_{\geq 0}$ such that $\gamma\leq \delta\le \eta$ and $B$ satisfy conditions~\eqref{eq_barr_11},~\eqref{eq_barr_21}, and the following condition
\begin{align}
    B(x) \leq \delta \implies \mathrm{inf}_{u\in U}\{B(f(x,u))\}\leq \delta.  \label{eq_barr_3}
\end{align} 
\end{defn}
The next result formulates the safe control synthesis problem as a search for CBCs~\cite{ADA-XX-JWG-PT:17}.

\begin{theorem}[Control synthesis via control barrier certificates]\label{thm:safebarrier2}
\label{cbcthm}
Consider a dtDS $\Sys = (\Xx, \Xx_0, U, f)$ and a set of unsafe states $\Xx_u \subseteq \Xx$. Suppose that there exists a control barrier certificate satisfying conditions~\eqref{eq_barr_11},~\eqref{eq_barr_21}, and~\eqref{eq_barr_3}. Then, there exists a map $K:\Xx\to U$ such that the system $\Sys$ with the feedback controller $\{u_t=K(x_t)\}_{t=0}^{\infty}$ is safe with respect to $\Xx_u$. 
\end{theorem}

\subsection{Monotone Dynamical Systems} In this subsection, we introduce a special class of discrete-time dynamical systems known as monotone systems. We begin by formally defining the notion of monotonicity. 

\begin{defn}[Monotone maps] Consider the sets $\mathcal{X}{\subseteq }\R^n$ and $\mathcal{Y}{\subseteq} \R^m$. Then the map $f:\mathcal{X}{\to} \mathcal{Y}$ is monotone if, for every $x,y\in \mathcal{X}$ such that $x\le y$, we have $f(x)\le f(y)$. 
\end{defn}

Using this notion of monotonicity, we define monotone dynamical systems.
\begin{defn}[Monotone dynamical systems]
A dtDS $\mathfrak{S}=(\mathcal{X},\mathcal{X}_0,U,f)$ is monotone if the state transition map $f:\Xx\times U\to \Xx$ is monotone. 
\end{defn}
For a differentiable transition map $f$ and convex sets $\Xx$ and $U$, the system $\mathfrak{S}$ is monotone if and only if $Df(x,u) \ge 0_{n\times(n+m)}$ for all $(x,u)\in \Xx\times U$~\cite{HLS:95}. Monotone systems preserve the standard partial order along their trajectories~\cite{HLS:95,angeli2003monotone}, resulting in highly regular behaviors that facilitate reasoning about dynamics via partial orders.

\paragraph{\textbf{Problem Formulation.}}

We consider the safety of a monotone dtDS $\Sys = (\Xx,\Xx_0,U,f)$ with an unknown transition map $f$. While $f$ is not explicitly known, we assume access to a simulator that returns $f(x, u)$ for any $x \in \Xx$ and $u \in U$. This report focuses on the following problem.

%
%


\begin{prob}[Safety Certificates for Unknown Monotone Systems]\label{prob:main}
Consider a monotone dtDS $\Sys = (\Xx, \Xx_0, U, f)$ with an unknown state-transition map $f$, and a set of unsafe states $\Xx_u \subseteq \Xx$. Our goal is to use a finite set of samples from a simulator of $f$ to:
\begin{enumerate}
    \item \textbf{(Safety verification):} Verify that $\Sys$ is safe with respect to $\Xx_u$ for all input $\{u_t\}_{t=0}^{\infty} \in U$.
    
    \item \textbf{(Safe control synthesis):} Design a feedback controller $K: \Xx \rightarrow U$ such that $\Sys$ with the feedback controller $\{u_t = K(x_t)\}_{t=0}^{\infty}$ is safe with respect to $\Xx_u$.
\end{enumerate}
\end{prob}

The monotonicity assumption in Problem~\ref{prob:main}, although made in the absence of an explicit model, is often justified by prior knowledge of the system's physical properties. Many physical processes naturally exhibit monotonicity, where increases in inputs or states lead to non-decreasing responses. This property is common in domains such as traffic networks~\cite{SC-MA:15}, power systems~\cite{park2020uniqueness}, biological processes~\cite{EDS:07}, and supply chains~\cite{yu2019equilibrium}.

\section{Monotone Neural Barrier Certificates}

We propose a data-driven framework that leverages system monotonicity to learn barrier functions using neural networks. This approach enables the construction of sound safety certificates from limited data, avoiding full system identification or exhaustive sampling. 

\subsection{Monotone Barrier Certificates}

In this subsection, we propose a suitable structure of barrier certificates that enables efficient data-driven safety verification and control synthesis for the monotone system $\Sys = (\Xx, \Xx_0, U, f)$.


\begin{theorem}[Monotone Barrier Certificates]\label{thm:monbar}
    Consider a monotone dtDs $\Sys :=(\Xx,\Xx_0,U,f)$ with an unsafe states $\Xx_u \subseteq \Xx$. Then, the following statements hold:
    \begin{enumerate}
        \item\label{p1} \textbf{(Safety verification).} $\Sys$ is safe with respect to $\Xx_u$, for every input $\{u_t\}_{t=0}^{\infty}\in U$ if  there exists a monotone robust barrier certificate satisfying~\eqref{eq_barr_11},\eqref{eq_barr_21}, and~\eqref{eq_barr_31}.
        \item\label{p2} \textbf{(Safe control synthesis).} there exists a map $K:\Xx\to U$ such that $\Sys$ with the feedback controller $\{u_{t} = K(x_t)\}_{t=0}^{\infty}$ is safe with respect to $\Xx_u$ if there exists a monotone control barrier certificate satisfying~\eqref{eq_barr_11},\eqref{eq_barr_21}, and~\eqref{eq_barr_3}.
    \end{enumerate}
\end{theorem}
\textit{Proof.} We refer to Appendix~\ref{app:monbar} for the proof.

For monotone systems, Theorem~\ref{thm:monbar} provides a sufficient condition for searching for robust (control) barrier certificates within the class of monotone barrier functions.
As will be shown in Section~\ref{sec:data}, the monotonicity structure of barrier certificates plays a crucial role in our framework, leading to significantly lower sample complexity and computational cost—particularly in high-dimensional systems.
This structure also enables restricting learning-based approximators to inherently monotone functions. Monotone neural networks thus provide a natural and expressive choice for approximating such certificates.

\paragraph{\textbf{Monotone Neural Networks.}} We consider a neural network $N$ with $k$ fully connected layers, where each layer $i$ is defined by a weight matrix $W_i$ and a bias vector $b_i$ of appropriate dimensions, followed by an activation function $\sigma$. The network represents a function $N: \mathbb{R}^{n_0} \to \mathbb{R}^{n_k}$. Given an input $u \in \mathbb{R}^{n_0}$, the network computes its output $y_k \in \mathbb{R}^{n_k}$ as follows:
\begin{align*}
y_0 &= u, \\
y_{i+1} &= \sigma(W_i y_i + b_i), \quad \text{for all } i \in \{0, 1, \ldots, k-1\}, \\
N(u) &= W_k y_k + b_k.
\end{align*}
Let $y_{i-1}$ and $y_i$ denote the input and output of the $i$-th layer, with activation function $\sigma:\real\to \real$. The network $N$ is monotone if (i) all weight matrices satisfy $W_i \ge 0_{n_i \times n_{i+1}}$ and (ii) $\sigma$ is non-decreasing. Monotone neural networks are universal approximators of monotone functions~\cite{daniels2010monotone}.

By Theorem~\ref{thm:monbar}, the search for robust (respectively, control) barrier certificates reduces to finding monotone neural networks that satisfy conditions~\eqref{eq_barr_11}, \eqref{eq_barr_21}, and~\eqref{eq_barr_31} (respectively,~\eqref{eq_barr_3}). However, formally verifying such certificates remains intractable due to the need to evaluate the network over a continuum of states. In the next section, we show how the system's and certificate's monotonicity enable safety verification and control synthesis from finite data.

\subsection{Monotone Barrier Certificates from Finite Data}\label{sec:data}

In this section, we exploit the system's and certificates' monotonicity to develop a data-driven framework for safety verification and control synthesis using only finite data, while ensuring formal correctness. We partition the state space $\Xx$ and the input space $U$ into hyper-rectangles $\{[\underline{x}^i, \overline{x}^i]\}_{i \in \mathcal{P}}$ and $\{[\underline{u}^i, \overline{u}^i]\}_{i \in \mathcal{S}}$, respectively, where $\mathcal{P}$ and $\mathcal{S}$ are finite index sets corresponding to these partitions. Given $\Xx_0 \subseteq \Xx$ and $\Xx_u \subseteq \Xx$, we define index sets $\mathcal{I}, \mathcal{U}$ as the minimal subsets of $\mathcal{P}$ such that
\begin{align}\label{eq:index}
\Xx_0 \subseteq \bigcup_{i \in \mathcal{I}} [\underline{x}^i, \overline{x}^i], \qquad
\Xx_u \subseteq \bigcup_{i \in \mathcal{U}} [\underline{x}^i, \overline{x}^i].
\end{align}
By construction, any $x \in \Xx_0$ satisfies $x \leq \overline{x}^{i}$ for some $i \in \mathcal{I}$, and any $x \in \Xx_u$ satisfies $\underline{x}^{i} \leq x$ for some $i \in \mathcal{U}$. For monotone systems with monotone barrier certificates, the points $\{\underline{x}^i\}_{i \in \mathcal{U}}$ and $\{\overline{x}^i\}_{i \in \mathcal{I}}$ act as \textit{worst-case} representatives. Thus, the search for robust barrier certificates satisfying~\eqref{eq_barr_11}-\eqref{eq_barr_31} can be conservatively reformulated as finding a monotone neural network $N:\Xx \to \mathbb{R}$ satisfying:
\begin{align}
    N(\overline{x}^i) &\le \gamma \quad &&\forall i \in \mathcal{I}, \label{eq:training_initial}\\
    N(\underline{x}^i) &> \eta \quad &&\forall i \in \mathcal{U}, \label{eq:training_unsafe}\\
    N(\underline{x}^i) &\leq \delta \quad\implies\quad \max_{j\in \mathcal{S}}N(f(\overline{x}^i, \overline{u}^j)) \leq \delta \quad &&\forall i\in \mathcal{P}. \label{eq:training_invariance}
\end{align}
A monotone neural network $N:\Xx\to \real$ satisfying~\eqref{eq:training_initial}, \eqref{eq:training_unsafe}, and~\eqref{eq:training_invariance} is called a \textit{monotone neural robust barrier certificate}. Similarly, for safe control synthesis, the search can be reformulated as finding a monotone neural network $N:\Xx \to \mathbb{R}$ satisfying~\eqref{eq:training_initial}, \eqref{eq:training_unsafe}, and
\begin{align}
    N(\underline{x}^i) &\leq \delta \quad\implies\quad \min_{j\in \mathcal{S}}N(f(\overline{x}^i, \underline{u}^j)) \leq \delta \quad &&\forall i\in \mathcal{P}. \label{eq:training_invariance-2}
\end{align}
A monotone neural network $N:\Xx\to \real$ that satisfies conditions~\eqref{eq:training_initial}, \eqref{eq:training_unsafe}, and~\eqref{eq:training_invariance} is referred to as a \textit{monotone neural control barrier certificate}. 
Conditions~\eqref{eq:training_initial}–\eqref{eq:training_invariance-2} reduce the verification of neural barrier certificates from the entire (continuous) state set to a finite set of carefully selected maximal and minimal points derived from the coverings of $\Xx$ and $U$.
The following result provides formal guarantees for safety verification and safe control synthesis using monotone neural barrier certificates. 
\begin{theorem}[Monotone Neural Barrier Certificates]\label{thm:guarantee}
    Consider a monotone dtDS $\Sys=(\Xx,\Xx_0,U,f)$ with an unknown state transition map $f$ and a set of unsafe states $\Xx_u$. Let $\{[\underline{x}^i, \overline{x}^i]\}_{i \in \mathcal{P}}$ be a family of hyper-rectangles that cover the state set $\Xx$, $\{[\underline{u}^i, \overline{u}^i]\}_{i \in \mathcal{S}}$ be a family of hyper-rectangles that cover the input set $U$, and the index sets $\mathcal{I},\mathcal{U}$ be the minimal subsets of $\mathcal{P}$ satisfying~\eqref{eq:index}. Then, 
    \begin{enumerate}
        \item\label{p11} \textbf{(Safety verification)} If there exists a monotone neural network $N:\Xx\to \real$ satisfying~\eqref{eq:training_initial},~\eqref{eq:training_unsafe}, and~\eqref{eq:training_invariance}, with $\gamma \leq \delta \leq \eta$, then $\Sys$ is safe with respect to $\Xx_u$, for every input $\{u_t\}_{t=0}^{\infty}\in U$; 
        \item\label{p22} \textbf{(Safe control synthesis)} If there exists a monotone neural network $N:\Xx\to \real$ satisfying~\eqref{eq:training_initial},~\eqref{eq:training_unsafe}, and~\eqref{eq:training_invariance-2}, with $\gamma \leq \delta \leq \eta$, then there exists a map $K:\Xx\to U$ such that $\Sys$  with the feedback controller $\{u_{t} = K(x_t)\}_{t=0}^{\infty}$ is safe with respect to $\Xx_u$.
    \end{enumerate}
\end{theorem}
\textit{Proof.} We refer to Appendix~\ref{app:guarantee} for the proof.


Theorem~\ref{thm:guarantee} highlights key advantages of our data-driven approach. It allows constructing safety certificates from arbitrary state space partitions, avoiding the curse of dimensionality in fine discretizations~\cite{anand2023formally} and the reliance on SMT solvers for neural network verification~\cite{edwards2024fossil}. By leveraging system monotonicity, it scales to high-dimensional systems and does not require knowledge of Lipschitz constants for the dynamics or barrier function.

\subsection{Training Monotone Barrier Certificates.}\label{sec:training}

In this section, we present an adaptive algorithm for training monotone neural robust (resp. control) barrier certificates. Let $\{[\underline{x}^i, \overline{x}^i]\}_{i \in \mathcal{P}}$ and $\{[\underline{u}^i, \overline{u}^i]\}_{i \in \mathcal{S}}$ be hyper-rectangular coverings of the state and input spaces, respectively, and $\mathcal{I}, \mathcal{U}$ the index sets defined in~\eqref{eq:index}.

For safety verification, we train a monotone neural network $N: \Xx \to \mathbb{R}$ as a robust barrier certificate by minimizing the composite mean-squared-error loss $L_{\mathrm{total}} := L_1 + L_2 + L_3$, where each term enforces a specific barrier certificate constraint:
\begin{align}
    L_1 &:= \mathrm{MSE}(N(\overline{x}^i), \gamma),
    && \text{for all } i \in \mathcal{I}, \label{eq:loss1} \\
    L_2 &:= \mathrm{MSE}(N(\underline{x}^i), \eta),
    && \text{for all } i \in \mathcal{U}, \label{eq:loss2} \\
    L_3 &:= \mathrm{MSE}\left( \min_{j \in \mathcal{S}} N(f(\overline{x}^i, \underline{u}^j)), \delta \right),
    && \text{for all } i \in \mathcal{P}. \label{eq:loss3}
\end{align}
Training uses gradient-based optimization on $L_{\mathrm{total}}$, with tunable parameters $\gamma, \delta, \eta \in \mathbb{R}_{>0}$ satisfying $\gamma < \delta < \eta$. The process stops once $N$ meets conditions~\eqref{eq:training_initial}, \eqref{eq:training_unsafe}, and~\eqref{eq:training_invariance}.

For safe control synthesis, we jointly train two monotone neural networks: $N: \mathcal{X} \to \mathbb{R}$, which serves as the control barrier certificate, and $K: \mathcal{X} \to \mathcal{U}$, which represents the feedback policy. The objective for both of these networks is the composite mean-squared-error loss $\widehat{L}_{\mathrm{total}} := L_1 + L_2 + L_4$, where $L_1$ and $L_2$ are defined in~\eqref{eq:loss1} and~\eqref{eq:loss2}, and $L_4$ is given by
\begin{align}
    L_4 &:= \mathrm{MSE}\left( N(f(\overline{x}^i, K(\overline{x}^i))), \delta \right), 
    && \mbox{ for all } i \in \mathcal{P}. \label{eq:loss4}
\end{align}
Training is performed jointly on $N$ and $K$ using gradient-based optimization on $\widehat{L}_{\mathrm{total}}$, with tunable parameters $\gamma, \delta, \eta \in \mathbb{R}_{>0}$ such that $\gamma \le \delta \le \eta$. It stops when $N$ satisfies~\eqref{eq:training_initial}, \eqref{eq:training_unsafe}, and the following condition:
\begin{align*}
    N(\underline{x}^i) \le \delta \quad \Longrightarrow \quad N(f(\overline{x}^i, K(\overline{x}^i))) \le \delta, \qquad \forall i \in \mathcal{P}.
\end{align*}

\section{Conclusions}
This work advances neurosymbolic verification by proposing a scalable framework for synthesizing safety controllers in high-dimensional \emph{monotone} systems using \emph{monotone neural barrier certificates}. Leveraging the systems' order-preserving structure, our approach reduces verification of monotone systems to local boundary checks. 
\bibliographystyle{abbrv}
\bibliography{ref,SJ,trivedi}

\newpage

\appendix

\section{Proof of Theorem~\ref{thm:monbar}}\label{app:monbar}

 Suppose that $B:\Xx\to \real$ is a monotone function satisfying conditions~\eqref{eq_barr_11},\eqref{eq_barr_21}, and~\eqref{eq_barr_31}. By definition~\ref{cbc_def}, $B:\Xx\to \real$ is a robust barrier certificate for $\Sys$ with respect to $\Xx_u$. Therefore, by Theorem~\ref{thm:safebarrier}, $\Sys$ is safe with respect to $\Xx_u$, for every input $\{u_t\}_{t=0}^{\infty}$.

Now we focus on the proof of Theorem~\ref{thm:monbar}~part~\ref{p2}. Suppose that $B:\Xx\to \real$ is a monotone barrier certificate satisfying~\eqref{eq_barr_11},\eqref{eq_barr_21}, and~\eqref{eq_barr_3}. Then, by Theorem~\ref{thm:safebarrier2}, there exists a map $K:\Xx\to \real$ such that the system $\Sys$ with the feedback controller $\{u_{t} = K(x_t)\}_{t=0}^{\infty}$ is safe with respect to $\Xx_u$.

\section{Proof of Theorem~\ref{thm:guarantee}}\label{app:guarantee}

Regarding part~\ref{p11}, our goal is to show that  the monotone neural network $N:\Xx\to \real$ satisfying conditions~\eqref{eq:training_initial},~\eqref{eq:training_unsafe}, and~\eqref{eq:training_invariance} is a robust barrier certificate for $\Sys$ with respect to the unsafe set $\Xx_u$. More specifically, our goal is to show that the monotone neural network $N:\Xx\to \real$ satisfies conditions~\eqref{eq_barr_11},~\eqref{eq_barr_21}, and~\eqref{eq_barr_31}.

 \begin{enumerate}
    \item\textbf{($N$ satisfies condition~\eqref{eq_barr_11})} Let $x\in \Xx_0$. Since $\Xx_0\subseteq \bigcup_{i\in \mathcal{I}}[\underline{x}^i,\overline{x}^i]$, there exists 
$k\in \mathcal{I}$ such that $x\le \overline{x}^{k}$. Since $N:\Xx\to \real$ satisfies condition~\eqref{eq:training_initial}, we have $N(\overline{x}^k)\le \gamma$. On the other hand, $N$ is a monotone neural network, thus $N(x)\le N(\overline{x}^k)\le \gamma$. This shows that the monotone neural network $N$ satisfies condition~\eqref{eq_barr_11}.
    
    \item\textbf{($N$ satisfies condition~\eqref{eq_barr_21})} Let $x\in \Xx_u$. Since $\Xx_u\subseteq \bigcup_{i\in \mathcal{U}}[\underline{x}^i,\overline{x}^i]$, there exists 
$j\in \mathcal{U}$ such that $\underline{x}^{j}\le x$. Since $N$ satisfies the condition~\eqref{eq:training_unsafe}, we have $\eta\le N(\underline{x}^j)$. Using the monotonicity of the neural network $N$, we get  $\eta\le N(\underline{x}^j)\le N(x)$. This means that the monotone neural network $N$ satisfies condition~\eqref{eq_barr_21}.
    
    \item\textbf{($N$ satisfies condition~\eqref{eq_barr_31})} Let $x\in \Xx$ be such that $N(x)\le \delta$. Since the family of hyper-rectangles $\{[\underline{x}^i,\overline{x}^i]\}_{i\in \mathcal{P}}$ covers the state set $\Xx$, there exists $k\in \mathcal{P}$ such that $x\in [\underline{x}^k,\overline{x}^k]$. Using the monotonicity of $N$, we get $N(\underline{x}^k) \le N(x) \le \delta$. Since $N$ satisfies condition~\eqref{eq:training_invariance}, we have $\max_{j\in \mathcal{S}} N(f(\overline{x}^k,\overline{u}^j)) \le \delta$. As a result, for every $u\in U$, we have
\begin{align*}
    N(f(x,u)) \le N(f(\overline{x}^k,u)) \le \max_{j\in \mathcal{S}} N(f(\overline{x}^k,\overline{u}^j)) \le \delta, 
\end{align*}
where the first inequality holds by the monotonicity of $N$ and $f$ and the fact that $x\le \overline{x}^k$, the second inequality holds by  monotonicity of $N$ and $f$ and the fact that $u\le \max_{j\in \mathcal{S}}\overline{u}^j$. This means that $N$ satisfies condition~\eqref{eq_barr_31}. 
\end{enumerate}
Therefore, the monotone neural network $N:\Xx\to \real$ satisfies conditions~\eqref{eq_barr_11},~\eqref{eq_barr_21}, and~\eqref{eq_barr_31}. The result then follows using Theorem~\eqref{thm:safebarrier}. This completes the proof of part~\ref{p11}.

Regarding part~\ref{p22}, our goal is to show that the monotone neural network $N:\Xx\to \real$ satisfying conditions~\eqref{eq:training_initial},~\eqref{eq:training_unsafe}, and~\eqref{eq:training_invariance-2} is a control barrier certificate for $\Sys$ with respect to $\Xx_u$. More specifically, our goal is to show that $N$ satisfies conditions~\eqref{eq_barr_11},~\eqref{eq_barr_21}, and~\eqref{eq_barr_3}.

 \begin{enumerate}
    \item\textbf{($N$ satisfies condition~\eqref{eq_barr_11})} Let $x\in \Xx_0$. Since $\Xx_0\subseteq \bigcup_{i\in \mathcal{I}}[\underline{x}^i,\overline{x}^i]$, there exists 
$k\in \mathcal{I}$ such that $x\le \overline{x}^{k}$. Since $N:\Xx\to \real$ satisfies condition~\eqref{eq:training_initial}, we have $N(\overline{x}^k)\le \gamma$. On the other hand, $N$ is a monotone neural network, thus $N(x)\le N(\overline{x}^k)\le \gamma$. This shows that the monotone neural network $N$ satisfies condition~\eqref{eq_barr_11}.
    
    \item\textbf{($N$ satisfies condition~\eqref{eq_barr_21})} Let $x\in \Xx_u$. Since $\Xx_u\subseteq \bigcup_{i\in \mathcal{U}}[\underline{x}^i,\overline{x}^i]$, there exists 
$j\in \mathcal{U}$ such that $\underline{x}^{j}\le x$. Since $N$ satisfies the condition~\eqref{eq:training_unsafe}, we have $\eta\le N(\underline{x}^j)$. Using the monotonicity of the neural network $N$, we get  $\eta\le N(\underline{x}^j)\le N(x)$. This means that the monotone neural network $N$ satisfies condition~\eqref{eq_barr_21}.
    
    \item\textbf{($N$ satisfies condition~\eqref{eq_barr_3})} Let $x\in \Xx$ be such that $N(x)\le \delta$. Since the family of hyper-rectangles $\{[\underline{x}^i,\overline{x}^i]\}_{i\in \mathcal{P}}$ covers the state set $\Xx$, there exists $k\in \mathcal{P}$ such that $x\in [\underline{x}^k,\overline{x}^k]$. Using the monotonicity of $N$, we get $N(\underline{x}^k) \le N(x) \le \delta$. Since $N$ satisfies condition~\eqref{eq:training_invariance-2}, we have $\min_{j\in \mathcal{S}} N(f(\overline{x}^k,\overline{u}^j)) \le \delta$. Since both $N$ and $f$ are monotone functions, for every $u\in U$, we have $N(f(x,u)) \le N(f(\overline{x}^k,u))$. By taking the infimum of both sides of this inequality, we obtain
\begin{align}\label{eq:inequality}
    \inf_{u\in U} N(f(x,u)) \le \inf_{u\in U} N(f(\overline{x}^k,u)) \le \min_{j\in \mathcal{S}} N(f(\overline{x}^k,\overline{u}^j)) \le \delta, 
\end{align}
where the second inequality holds using the fact that $\{[\underline{u}^i,\overline{u}^i]\}_{i\in \mathcal{S}}$ is a hyper-rectangular cover for $U$. 
The inequality~\eqref{eq:inequality} implies that $N$ satisfies condition~\eqref{eq_barr_3}. 
\end{enumerate}
Therefore, the monotone neural network $N:\Xx\to \real$ satisfies conditions~\eqref{eq_barr_11},~\eqref{eq_barr_21}, and~\eqref{eq_barr_3}. The result then follows using Theorem~\eqref{thm:safebarrier2}. This completes the proof of part~\ref{p22}.

\end{document}